\documentstyle[12pt,epsfig,aj_pt4]{article}
\tighten
\lefthead{Gallart et al.}
\righthead{The SFH of Leo I}

\begin{document}

\title{The Star Formation History of the Local Group dwarf galaxy Leo~I\altaffilmark{1}}

\author{Carme Gallart\altaffilmark{2,3}, Wendy L. Freedman\altaffilmark{2},   Antonio Aparicio\altaffilmark{4}, Giampaolo Bertelli\altaffilmark{5, 6}, Cesare Chiosi\altaffilmark{5}}

Subject Headings:  galaxies: individual (Leo I); galaxies: evolution; galaxies: stellar content; galaxies: photometry; stars: Hertzsprung-Russell (HR-diagram).

\altaffiltext{1}{Based on observations with the NASA/ESA {\it Hubble Space Telescope} obtained at the Space Telescope Science Institute, which is operated by AURA, Inc., under NASA contract NAS 5-26555.}
\altaffiltext{2}{Observatories of the Carnegie Institution of Washington, 813 Santa Barbara St., Pasadena, California 91101, USA}
\altaffiltext{3}{Yale Andes Prize Fellow (Yale University and Universidad de Chile. Casilla 36-D. Las Condes. Santiago. Chile.)} 
\altaffiltext{4}{Instituto de Astrof\'\i sica de Canarias, E-38200 La Laguna, Canary Islands, Spain} 
\altaffiltext{5}{Dipartimento di Astronomia dell'Universit\`a di Padova, Vicolo dell'Osservatorio 5, I-35122- Padova, Italy}
\altaffiltext{6}{National Council of Research, CNR-GNA, Rome, Italy}

\begin{abstract}

We present a quantitative analysis of the star formation history (SFH) of the Local Group dSph galaxy Leo I, from the information in its {\it Hubble Space Telescope} $[(V-I), I]$ color-magnitude diagram (CMD). It reaches the level of the oldest main-sequence turnoffs, and this allows us to retrieve the SFH in considerable detail. The method we use is based in comparing, via synthetic CMDs, the expected distribution of stars in the CMD for different evolutionary scenarios, with the observed distribution. We consider the SFH to be composed by the SFR(t), the chemical enrichment law Z(t), the initial mass function IMF, and a function $\beta(f,q)$, controlling the fraction $f$ and mass ratio distribution $q$ of binary stars. We analyze a set of $\simeq$ 50 combinations of four Z(t), three IMF and more than four $\beta(f,q)$. For each of them, the best SFR(t) is searched for among $\simeq 6 \times 10^7$ models. The comparison between the observed CMD and the model CMDs is done through $\chi^2_{\nu}$ minimization of the differences in the number of stars in a set of regions of the CMD, chosen to sample stars of different ages or in specific stellar evolutionary phases. We empirically determine the range of $\chi^2_{\nu}$ values that indicate acceptable models for our set of data using tests with models with known SFHs.

Our solution for the SFH of Leo I defines a minimum of $\chi^2_{\nu}$ in a well defined position of the parameter space, and the derived SFR(t) is robust, in the sense that its main characteristics are unchanged for different combinations of the remaining parameters. However, only a narrow range of assumptions for Z(t), IMF and $\beta(f,q)$ result in a good agreement between the data and the models, namely: Z=0.0004, a Kroupa et al. (1993) IMF or slightly steeper, and a relatively large fraction of binary stars, with $f=0.3-0.6$, $q>0.6$ and approximately flat IMF for the secondaries, or particular combinations of these parameters that would produce a like fraction of similar mass binaries. Most star formation activity (70\% to 80\%) occurred between 7 and 1 Gyr ago. At 1 Gyr ago, it abruptly dropped to a negligible value, but seems to have been active until at least $\simeq 300 $ million years ago. Our results don't unambiguously answer the question of whether Leo I began forming stars around 15 Gyr ago, but it appears that the amount of this star formation, if existing at all, would be small.   

\end{abstract}

\section{Introduction: the SFHs of the nearest dSph galaxies} \label{intro}

In the last few years, a number of very deep, wide field color-magnitude diagrams (CMDs) have spectacularly revealed the stellar populations of some of the nearest dSph galaxies (Mighell \& Rich 1996; Smecker--Hane et al. 1996; Stetson 1997; Stetson et al. 1998;  Gallart et al. 1999a -Paper I-). The picture offered by these CMDs on the evolution of dSph galaxies is a complex and varied one, quite different from the early idea (Baade 1963) that dSph galaxies basically consisted of pure Population II stars (although differences from the CMD of globular clusters were early recognized: Baade 1963; Baade \& Swope 1961).

Comparison with globular clusters has been an useful tool, exploited until the present, for the analysis of the stellar populations in dSph: the presence of an horizontal-branch (HB) and RR Lyrae stars indicates the presence of a ``globular cluster age'' population; the position of the red-giant branch (RGB) and its width have been used to obtain an estimate of the metallicity and the metallicity dispersion respectively (Sandage 1982; Da Costa \& Armandroff 1990; Lee, Freedman \& Madore 1993c). The presence of stars younger than ``globular cluster age'' was suggested by the discovery of Carbon stars in dSph (Westerlund 1979; Aaronson \& Mould 1980; Mould et al. 1982; Frogel et al. 1982; Azzopardi, Lequeux \& Westerlund 1986), and bright AGB stars in general became a diagnostic for the presence of an intermediate age population (e.g. Elston \& Silva 1992; Freedman 1992; Lee, Freedman \& Madore 1993b; Davidge 1994; but see also Mart\'\i nez-Delgado \& Aparicio 1997). Stellar evolution theory became increasingly important to interpret the CMDs of dSph galaxies when they got deep enough to reach the main sequence (MS), and the intermediate-age and old turn-offs were revealed (e.g. in Carina, Mould \& Aaronson 1983; Mighell 1990a). Comparison of the position of the stars in the CMD with the position of theoretical isochrones of given age allows an estimate of the range of ages present in the stellar population. To estimate the fraction of stars formed in different star formation episodes, it is necessary to compare the number of stars as a function of luminosity and/or color with the numbers predicted by stellar evolution. Studies comparing observed and theoretical luminosity functions have been undertaken, among others, by Mould \& Aaronson (1983), Eskridge (1987), Mighell (1990b), Mighell \& Butcher (1992), Grillmair et al. (1998) and Hurley-Keller, Mateo \& Nemec (1998).  

In Paper I we used an isochrone fitting technique to study the stellar populations in Leo I from deep $HST$ F555W ($V$) and F814W ($I$) observations of a central field in this galaxy. The resulting CMD reaches $I \simeq 26$ and reveals the oldest  $\simeq 10-15$ Gyr old turnoffs, even though a HB is not obvious in the CMD. Given the low metallicity of the galaxy, we concluded that the absence of a conspicuous HB likely indicates that the first substantial star formation in the galaxy may have been somehow delayed in comparison with the other dSph satellites of the Milky Way. The structure of the red-clump (RC) of core He-burning stars is consistent with the large intermediate--age population inferred from the MS and the subgiant region. In spite of the lack of gas in Leo~I, the CMD clearly shows star formation continuing until 1 Gyr ago and possibly until a few hundred Myrs ago in the central part of the galaxy.

In this paper, we take advantage of the detailed predictions of stellar evolution theory on both the positions and numbers of stars across the CMD, to produce synthetic CMDs that are compared with the observed CMD to explore a wide range of parameters defining the star formation history (SFH) of Leo~I. This approach has been discussed by Gallart et al. (1996b) and by Aparicio, Gallart \& Bertelli (1997a,b), and allows the SFH to be obtained from the information in the CMD. It is similar also to the methods explored theoretically by different authors (e.g. Dolphin 1997; Hern\'andez, Valls-Gabaud \& Gilmore 1998; Ng 1998), but it is applied here to obtain the SFH from real data, and therefore, it takes into account the complications derived from actual observational errors, and the not-perfect stellar evolution theory, as discussed in Section~\ref{boxes} (see also Hurley-Keller et al. 1998 and Tolstoy \& Saha 1996). In the current paper, we explore the sensitivity of this approach to a number of parameters including, besides the star formation rate as a function of time SFR(t), the metallicity Z(t), the initial mass function (IMF), and the characteristics of the binary star population. We also undertake a number of tests with input models with known SFH in order to assess the reliability of this method. The Leo~I data are much deeper than the data we have previously analyzed using this technique and, therefore, allow much more detailed information on the SFH to be obtained for this galaxy. Data of similar or even better quality are available (or can be obtained) for all the satellites of the Milky Way, and therefore, by using the approach presented here, we expect to obtain their SFHs with at least this level of detail.

This paper is organized as follows: in Section 2, we summarize the characteristics of the Leo I data; in Section 3 we give an overview of our method to retrieve the SFH. In Section 4 we discuss the ingredients of our models, namely the input stellar evolution models, and the parameter space we will explore for: the IMF, the chemical enrichment, and the characteristics of the binary star population. In Section 5, we describe the details of the process of retrieving the SFH of Leo I. In Sections 6 and 7, our results are described and discussed. We describe the tests performed to characterize the accuracy of our method and the derived solutions in Appendix A.

\section{The Leo I data} \label{data}

WFPC2 {\it HST V} (F555W) and $I$ (F814W) data in one $2.6\arcmin \times 2.6\arcmin$ field in the center of Leo I were obtained on March 5, 1994. Three deep exposures with both F555W and F814W filters were taken (1990 sec and 1600 sec  respectively). To ensure that the brightest stars were not saturated, one shallow exposure with each filter (350 sec and 300 sec) was also obtained. Photometry of the stars in Leo I was obtained using the set of DAOPHOT II/ALLFRAME programs developed by Stetson (1994), and the final photometry on the Johnson-Cousins system was calibrated using the ground-based photometry obtained by Lee et al. (1993a).

In Figure~\ref{leoi234} we present the $[(V-I)_0, M_I]$ CMD of Leo I based on the three WF chips. It contains a total of 28,000 stars with small photometric errors ($\sigma \le 0.2$ chi$\le$ 1.6 and -0.5 $\le$ sharp $\le$ 0.5, as given by ALLFRAME). For a complete  description of the data reduction and a discussion of the features present in the CMD, see Paper I.


\section{Simulating the observational errors in the synthetic CMDs} \label{errors}

The simulation of the observational errors in the synthetic CMDs has been done on a star-by star basis, using an empirical approach that doesn't make  any assumption about the nature of the errors or about their propagation. In essence, our strategy, based on artificial star tests, is the following: as the artificial star sample, we use a synthetic CMD similar to those that will be used in the derivation of the SFH. For each of these stars, we obtain an error in magnitude and color from artificial star tests. Since the synthetic CMD is densely populated, we have a large number of stars in each small interval of magnitude and color, and therefore, an statistically significant sampling of the errors across the CMD. Then, to simulate the errors in any other synthetic CMD, we apply to each of its stars the errors of an artificial star of similar magnitude and color, randomly chosen among those in a small magnitude and color interval in the CMD. Figure~\ref{colores} shows an example of (a) a synthetic CMD and (b) the corresponding model CMD after the simulation of the observational errors.


 The whole process has been performed in a similar way as described in Aparicio \& Gallart (1995) and Gallart et al. (1996a,b). In these papers, a detailed description of the observational errors was presented, in terms of crowding factors and magnitude shifts and errors as a function of both color and magnitude. Here, besides the fact that the procedure has been adapted to the current reduction of the data done with ALLFRAME, the main difference with the strategy followed in the above references is a) the distribution of the artificial stars in the CMD, using magnitudes and colors of the stars of a synthetic CMD as discussed above, and b) the optimization of the distribution of artificial stars in the real frames. We briefly comment on this last point below. 

Since all we require for artificial stars to represent the actual observational errors is that they don't interact with other artificial stars, but only with real stars, the distance between them on the image of the galaxy must be such that the effect of the wings of the neighboring artificial stars is negligible. Because the PSF is scaled to each star using only the information on the fitting radius (2 pixels here), if the artificial stars were (PSF radius + fitting radius +1) pixels apart (this is 17 pixels for the WF chips) there wouldn't be any interaction between them. However, to optimize the computer use, we made several tests to determine how closely the artificial stars can be placed in the image without introducing spurious overcrowding effects in their photometry. We found that 11 pixels (except for the brightest stars which were added with extra spacing) was a safe value that allows to add a sufficient number of stars to each galaxy image in order to reduce the number of crowding tests to a manageable value. Therefore, a grid with artificial stars spaced 11 pixels is displaced randomly for each test to assure a dense sampling of the observational effects all over the image of the galaxy.

\section{Retrieving the star formation history: comparison of model and observed CMDs} \label{method}

Our goal is to reconstruct the SFH of Leo I from the information in its CMD. The method we are using is based on the comparison of the distribution of stars in the observed CMD with that in model CMDs resulting from a large number of possible SFHs. We consider the SFH to be composed by the SFR(t), the chemical enrichment law Z(t), the initial mass function IMF, and a function $\beta(f,q)$, controlling the fraction $f$ and mass ratio distribution $q$ of binary stars. We analyze large sets of models with different combinations of Z(t), IMF and $\beta(f,q)$, following an approach similar to that of Aparicio et al. (1997b) in the case of the dwarf galaxy LGS~3: only one model CMD, with constant SFR(t) has been computed for each set of IMF, Z(t) and $\beta(f,q)$. Then, this model has been divided up into a number of age intervals, or step functions. Arbitrary SFR(t) functions can then be defined as linear combinations of the partial models. Once the number of stars with which each partial model populates each of the regions defined in the CMD  (see Figure~\ref{cajas6}) is known, the number of stars that populate that region for a given SFR(t) can be directly obtained as a linear combination of the form:
$$ N_{s,j} = \sum_{i=1}^{m} a_i N_{i,j}, \eqno (1)$$
where  $N_{s,j}$ is the number of stars in region $j$ for the combined (synthetic) model CMD; $N_{i,j}$ is the same number corresponding to partial model $i$, and $a_i$ are the coefficients that indicate the relative star formation rate in the interval of time represented by each partial model; $m$ is the number of partial models used. The $N_{s,j}$ values are all we need to compare the models with observations. An arbitrarily large number of models with different SFR(t) can be calculated by just changing $a_i$.    

The comparison between the observed CMD and the model CMDs is done through $\chi^2_{\nu}$ minimization of the differences in the number of stars in a set of regions of the CMD, chosen to sample stars of different ages or in specific stellar evolutionary phases. For each set of Z(t), IMF and $\beta(f,q)$, the minimum of the $\chi^2_{\nu}$ function, which will provide the best fitting SFR(t) for that particular set of functions, is searched for.  We calculate the reduced chi-squared, $\chi^2_{\nu}$, of the number of stars in the $r$ regions of the observed and each model CMD, as: 

$$\chi^2_{\nu}= \frac{1}{\nu}\sum_{j=1}^{r} \frac {[(l \sum_{i=1}^{m} a_i N_{i,j}) - N_{O,j}]^2} {N_{O,j}} \eqno (2)$$

\noindent where l is a normalization factor to set the total number of stars inside the regions used equal to the number of stars in the corresponding regions of the observed CMD, $N_{O,j}$ is the number of stars in region $j$ in the observed CMD, and $\nu$ is the number of degrees of freedom  (which is equal to the number of boxes minus 1). Whether the best fitting SFR(t) is a good representation of the data is indicated by the value of $\chi^2_{\nu}$ itself (see Section~\ref{cutoff}). 


Figure~\ref{colores} shows a synthetic CMD made up of a set of partial models. Note how stars of different ages occupy different regions of the CMD. There is a sequence of ages from bright (younger) to faint (older) stars, along the MS, subgiant branch and even although more entangled, in the RC/HB. These age sequences are better defined in the synthetic CMD, prior to the simulation of the errors (Figure~\ref{colores}a) and get blurred, as expected, when the errors are simulated  (Figure~\ref{colores}b). 

\vskip 0.8 true cm

\section{The parameter space} \label{param}

As discussed in the last section, in addition to the SFR(t), the characteristics of the stellar populations of a galaxy are defined mainly by the chemical enrichment law Z(t), the IMF, and other details of the star formation process like the characteristics of the binary star population $\beta(f,q)$. We will use estimates of the IMF from studies in the solar neighborhood and in other nearby galaxies. Z(t) is strongly dependent on both the
SFH and global characteristics of the galaxy that might determine the
SFH itself, like the mass and the interaction with the intergalactic
medium. We will use the hints provided by the morphology of the CMD to
put some initial constraints on Z(t), but we will explore a relatively
wide space for it as well as for $\beta(f,q)$. No assumptions will be made about the SFR(t), which will be obtained by minimization of $\chi^2_{\nu}$ for each choice of the Z(t), IMF and $\beta(f,q)$.

In general, we will adopt the distance to Leo I obtained by Lee et al. (1993a), $(m-M)_0=22.18 \pm 0.1$, but we will also carry on a set of tests taking into account the error in the distance modulus. The fact that we don't find any solution for values of $(m-M)_0$ at the extremes of the error interval (see Section~\ref{dist}) will support a) the adopted value for the distance modulus (at least relative the stellar evolution models used), and ii) the uniqueness of our solution, since this result implies that no combination of the model parameters can compensate distance errors. 

\subsection{Stellar evolutionary models}

The Padova library of stellar evolutionary tracks has been used, and in particular, the tracks calculated for the following set of chemical compositions: [$Y$=0.230, $Z$=0.0001] (Girardi et al. 1996), [$Y$=0.230, $Z$=0.0004] (Fagotto et al. 1994a), and [$Y$=0.240, $Z$=0.004] (Fagotto et al.  1994b). These are the most complete set of stellar evolutionary models currently available. They include a wide range of metallicities, masses and ages (from $4\times 10^6$ to $16\times10^9$ years), and the calculation of late stages of evolution, beyond the tip of the RGB, through the HB and AGB phase. A reliable semi-empirical conversion from theoretical to observational plane is also provided. For a summary of the properties of these models, see Bertelli et al. (1994). 

Since the reliability of our results on the SFH depends critically on that of the stellar evolution models in which they are based, we will briefly summarize the current main caveats of stellar evolution models in the mass range of interest here, namely 0.5-2.0 M$_{\odot}$, and the solutions adopted in the set of models used. Opacities, nuclear reaction rates and equation of state are not a problem because all modern models are calculated using the state of the art of these important physical ingredients. In fact, small changes in the equation of state with respect the precedent Padova models of higher Z were introduced by Girardi et al. (1996) in the computation of the Z=0.0001 models, and they concluded that the tracks don't change significantly. Problems still remain with mixing and mass loss. Here, the overshoot parameter has been callibrated from the analysis of CMD and luminosity functions of open clusters (Aparicio et al. 1990, Bertelli, Bressan \& Chiosi 1992; Carraro et al. 1993), and mass loss along the RGB is parameterized using the empirical formulation by Reimers (1975), with a value of the mass loss parameter callibrated using globular clusters CMDs. Finally, to convert from theoretical to observational plane, which introduces some uncertainties, in particular for the coolest stars, we have adopted the spectral energy distributions in the library of stellar spectra by Kurucz (1992, private communication) and empirical ones for the coolest stars (see Bertelli et al. 1994).       

How these uncertainties may affect the predicted distribution of stars in the CMD? First, it is worth noting that the different parameters intervening in the models have been callibrated using simple systems such as stellar clusters with considerable success. Therefore, it seems reasonable to assume that they are 'right' when analizing a complex stellar population like that of Leo I, in order to derive more global information like the SFH. In other words, the effect on the distribution of stars in the CMD will be much more affected by the assumed SFH than by the details of stellar evolution, which in this context may be considered negligible.  Second, the modelling gets more critical for advanced stellar evolutionary phases, such as the RGB and the HB. In our comparison between the observed and the model CMDs, the main weigth (13 out of 18 boxes in the CMD, see Section~\ref{boxes}) corresponds to the main sequence, the subgiant branch and low RGB -below the RC. Third, by using larger boxes in the areas of the CMD where stellar evolution or the transformation to the observational plane is more uncertain, we are giving less weight to the information in these areas (see also Section~\ref{boxes}).   

In order to quantify how much the uncertainties in stellar evolution may be affecting our conclusions, it would be useful to perform a parallel analysis of a given stellar population using two different sets of stellar evolution models. At the present time we don't have the necessary information on another set of models to do such parallel analysis, but work in that direction is in progress. For the moment, we will just mention  our comparison of the Padova (Bertelli et al. 1994) and Yale (Demarque et al. 1996) isochrones in Paper I. We noted a general good agreement between both sets, except for some differences regarding the shape of the RGB, and slight differences in the luminosity of intermediate-age subgiant branches. 



\subsection{The IMF} \label{imf1}

Following Scalo's (1998) suggestion to perform ``every calculation using at least two disparate choices of IMF'' and ``do not assume that some {\it best IMF} is known'', we computed three sets of models with three different IMFs with a common slope $m^{-1.3}$ for $0.08\le m < 0.5$ as in Kroupa,  Tout \& Gilmore (1993) for the disk of the Milky Way, and three different slopes for the upper mass interval, bracketing those of Kroupa et al.: 

a) $m^{-3.2}$  for $0.5\le m <  30.0 $

b) $m^{-2.2}$  for $0.5\le m < 1.0$; $m^{-2.7}$  for $1.0\le m < 30.0$ (Kroupa et al. 1993)

c) $m^{-2.0}$  for $0.5\le m <  30.0$

The stars that we observe in the CMD have masses above $M\simeq 0.6 M\odot$ and therefore, the counts in the CMD do not depend on the slope of the lower mass segment, although it does affect the zero point of the SFR(t) and hence the total mass in stars. The maximum mass considered is 30 $M_{\odot}$. 

\subsection{The enrichment law}

In Paper I we discussed possible limits on the Leo I metallicity and metallicity dispersion based on comparison with isochrones, and taking into account the dispersion in age of the stars in the galaxy. We concluded that either a) the width of the Leo~I RGB can be accounted for by the dispersion in stellar ages, and therefore, the metallicity dispersion could be negligible or b) considering the variation in color of the isochrones depending on both age and Z, the maximum range of metallicity could be $0.0001\le Z \le 0.001$.

We considered both possibilities in our simulations, and we calculated models with the following four metallicity laws (see Figure~\ref{zlaws}):

a) Constant Z=0.0001

b) Constant Z=0.0004

c) Constant Z=0.0006

d) Z increasing with time from Z=0.0001 to Z=0.0008 as shown in Figure~\ref{zlaws} (solid thick line), and with a Gaussian dispersion of $\sigma_{Z}=0.0001$ at each age. Z(t) has been determined by choosing the metallicity that would produce an isochrone centered in the observed RGB at each age.



\subsection{The binary fraction $\beta(f,q)$} \label{secbin}

Although the fraction of stars in binary systems and the mass distribution of binary companions is not yet known with certainty, existing studies point to a relatively large fraction of binary stars, both in the Galactic field and dSph galaxies (e.g. Halbwachs 1986; Duquennoy \& Mayor 1991; Kroupa, Gilmore \& Tout 1991; Olszewski, Pryor \& Armandroff 1996) and in clusters (e.g. Mermilliod \& Mayor 1989; Aparicio et al. 1990; Bolte 1992; Kroupa \& Tout 1992; Yan \& Mateo 1994; Rubenstein \& Baylin 1997; Elson et al. 1998). 

In their spectroscopic study of multiplicity among solar type stars in the solar neighborhood, Duquennoy \& Mayor (1991) found  that about 40\% of the ``stars'' in their sample are multiple systems and that the secondary mass distribution is similar to the Kroupa et al. (1990) IMF (but see also Mazeh et al. 1992). Kroupa et al. (1991) suggest that at least 50\% of the systems with primaries less massive than the Sun are composed of two stars with independent masses chosen from the same mass function. In relation to dSph galaxies, Olszewski et al. (1996) estimated that the binary frequency for stars near the turnoff in Draco and UMi is 3--5 times that found for the solar neighborhood.  

Studies of clusters indicate that, for more massive stars, approximately equal-mass companions are common. In a number of clusters, a `second sequence' about 0.75 mag brighter than the single-star MS is discernible. It runs parallel to the MS and must be a result of systems having a mass ratio $q$ greater than about 0.6 (Hurley \& Tout 1998, and references therein). Using different approaches, both Kroupa \& Tout (1992) and Elson et al. (1998) reach the conclusion that stars above $1 M_{\odot}$ appear more likely to have similar mass components than lower mass systems. In particular, in the case of the open cluster Praesepe, Kroupa \& Tout (1992) find that, for the more massive stars of their sample, the proportion of equal-mass systems is $\simeq 30 \%$ if $f$ among all stars is taken to be 1 and rises to  $\simeq 50 \%$ for $f\simeq 0.4$ For NGC1818, Elson et al. (1998) find that $f$ for the range of single-star masses 2.0--5.5 $M_{\odot}$ is $\simeq 20 \%$ (and larger in the core of the cluster), and that the majority of binary systems in their sample must consist of pairs with $m_2\ge0.7m_1$ (since they could not distinguish binaries with lower $q$, from single stars). 

We cannot uniquely determine the binary fraction, but we are interested in exploring the consequences of the presence of binaries with properties similar to those observed locally, on the CMD of Leo I. Since only for mass ratios $q$ close to unity (i.e. when the two components have similar masses) will the secondary have a substantial effect on the combined luminosity of the binary, in our models we have considered most mass ratios in the interval $0.6\le q \le 1.0$, and the masses of the secondary stars are drawn from the same IMF (see below). In some additional tests, we used different $q$ intervals and/or the secondary stars drawn from a flat IMF. We considered 5 different values of $f$: 0, 0.1, 0.3, 0.6 and 0.9. The main effect of the presence of binaries in a CMD like the one of Leo I is the increase in the number of stars in the subgiant-branch area (boxes 7-10 in Figure~\ref{cajas6}), which we cannot populate as much as in the observations otherwise. This underpopulation of the subgiant branch could be alternatively explained if our models were underestimating the lifetime of the stars in the subgiant branch with respect
the MS lifetime. However, since the models are relatively well determined in these phases, the binary star alternative seems to be more pausible. Binaries also produce the well known shift in the main position of the MS (to the red a to brighter magnitudes), but this affects little our star counts due to the shape and size of the boxes in the MS. The effects of binary stars on the CMD don't mimic those expected from an increased metallicity dispersion, or a shift in the distance modulus.

To generate the binary stars in the synthetic CMDs the following procedure is used. For each star, a random numbers generator is used to decide whether, according to $f$, it is or is not a member of a binary system. If it is, the companion star is calculated with the restrictions that its age and metallicity are the same as those of the former star. Its mass is obtained from the same IMF as for the single stars (or a flat IMF) with the restriction that it must be in the interval imposed by the limits of $q$.
     
\section{Searching the SFH of Leo~I}\label{search}

\subsection{Calculating the model CMDs} 

\noindent The model CMDs are calculated using a recent version of our synthetic CMD simulator. The core of this program, and its most distinctive feature, is the interpolation that is performed on the stellar evolutionary tracks of fixed mass and metallicity to determine the precise location in the CMD of stars of any age, mass and metallicity, with a resulting smooth distribution of stars following given SFR(t), IMF and Z(t). At the low metallicities of dwarf galaxies, the interpolation in metallicity is particularly important to explore (and match) the enrichment history of each galaxy.     

We have calculated a synthetic CMD with 100000 stars for each combination of the parameters defining Z(t), the IMF and $\beta(f,q)$, as discussed in Section~\ref{param}. The observational errors have been simulated as discussed in Section~\ref{errors}, using the artificial stars in each of the three WF chips. As a result, the final model CMDs have 300000 stars. 

\subsection{Defining the grid of regions in the CMD}\label{boxes}

Our approach to compare the models with the observed CMD involves a particular 2-dimensional binning of the stars as a function of their position in the CMD. This is a generalization of the luminosity function method, since we are taking into account information on the color of the stars in addition to their magnitude. Other theoretical approaches (Tolstoy \& Saha 1996; Dolphin 1997; Hern\'andez et al. 1998; Ng 1998) prefer a point-to-point approach to the comparison. However, the binning approach permits addressing a) the uncertainties in certain phases of stellar evolution, and in the transformation from the theoretical plane (${\rm T}_{\rm eff}$, L) to the observational CMD (which affect the positions of the stars in the CMD more than their numbers), and b) the difficulty of a ``perfect'' simulation of the observational errors, even using fully empirical approaches as in this paper.  Both issues imply some uncertainty on the position of the individual stars in the CMD. By using a two-dimensional binning, we are giving less weight to individual positions and more to the star counts in relatively large areas of the CMD, averaging out small scale fluctuations. The disadvantages of this approach are that a) it uses less information from the CMD than a point-to-point approach and, as a consequence, we may be able to recover the SFH in {\it apparently} less detail and b) the choice of age intervals determines the resolution of the SFH, (i.e., by choosing certain intervals of age we are deciding {\it a priori} to get the resulting SFR averaged over them). We tested the sensitivity of the derived SFH to these choices by analyzing CMDs obtained with known SFHs, in similar ways as described in Appendix A.  

Figure~\ref{cajas6} displays the boxes adopted in the CMD in this study to compare the distribution of the stars in the data and in the models. They have been chosen by taking into account the different stellar evolutionary phases present in the CMD. Their shapes are intended to separate stars of different ages by, for example, following the slope of the isochrones on the MS or the subgiant branch;  their sizes take into account the uncertainties in stellar evolution and the photometric errors. This final set of boxes is the result of an iterative procedure in which we ran a number of tests against known SFHs (as in Appendix~A) that indicated the usefulness of adding certain boxes and eliminating those that were redundant. We also checked the effects of changing the size of the boxes. Apart from minor differences, the results of these tests indicated that the results don't greatly vary for sets of boxes reasonably distributed in the CMD.

\subsection{Finding the minimum of the chi-squared function}\label{minchi}

 The chi-squared function (eq. 2 in Section~\ref{method}), is a function of $m$ independent variables, the SFR(t) coefficients $a_i, (i=1,...,m$,  with $m$ being the number of partial models or intervals of age) used to construct the synthetic CMDs. For a given set of IMF, Z(t), $\beta(f,q)$ and distance modulus,  the best SFR(t) will be defined by the set of coefficients $a_i$ that minimize the chi-squared function.  

The $m$-dimensional hyper-surface we are dealing with is, in general, quite complex, and in order to find the absolute minimum, we evaluate the function in a grid of values of the coefficients $a_i$. If $m$ is the number of partial models and $k$ the number of possible values of the coefficients $a_i$, the number of models that need to be evaluated following this approach is $k^m$. Computational limitations restrict the number of partial models $m$ (and therefore the temporal resolution) and the number of coefficients $k$ (and therefore the range of amplitudes of the SFR(t) to be tested)\footnote{ We have also tried an optimized search for the minimum of the hyper-surface using a downhill simplex method in multi-dimensions (Nelder \& Mead 1965; Press et al. 1986), with which we have obtained results consistent with the grid search}. 

 Given a combination of IMF, Z(t) and $\beta(f,q)$, a set of partial models is defined by a set of time steps $t_i$. After some testing, we found that a good compromise was to use 10 time intervals, (hence $m=10$ partial models) and $k=6$ possible values for the $a_i$ coefficients (0 to 1 in steps of 0.2). This means $6^{10}\simeq 6\times10^7$ models for each combination of IMF, Z(t) and $\beta(f,q)$. This gives us time intervals  $\Delta t \simeq 1$ Gyr (which is about the intrinsic age resolution in the CMD) and a regular sampling of the amplitude of the SFR(t) in relatively small steps of 0.2.

The 10 partial models are delineated by the age steps 15.0, 12.0, 9.4, 7.2, 5.4, 3.9, 2.7, 1.8, 1.1, 0.5 and 0.2 Gyr\footnote{We assumed that no star formation has taken place in Leo I in the last 0.2 Gyr.}. Note that we have chosen the interval of time defining each partial model to increase toward older ages, in correspondence both with the packing of the older stars in the CMD, and with the higher photometric errors at fainter magnitudes. By choosing these intervals of age for the partial models we are introducing an upper limit for the resolution in age of the SFR(t), since the values of the SFR(t) that we will obtain will be, in fact, an average over the period of time encompassed by each partial model.

As discussed in Section~\ref{method}, we calculate $\chi^2_{\nu}$ of the number of stars in each box  in the observed and each of the $k^m$ models as an indicator of the goodness of fit. One of these combinations will produce a minimum value of $\chi^2_{\nu}$, that we note $\chi^2_{\nu, min}$. In practice, there are a number of $a_i$ sets that produce very similar values of $\chi^2_{\nu}$ near the minimum. This is mainly due to the coarse sampling of the parameter space that allows us to approach the actual minimum, but not to find its precise position (since our grid may not have a node on it). We obtain the best SFR(t) as the average of a set of $a_i$ coefficients. We keep for the average all the models with $\chi^2_{\nu} \le \chi^2_{\nu, lim}$ with  $\chi^2_{\nu, lim}= \chi^2_{\nu, min} + 1.0$\footnote {This is equivalent to keeping a region in the space of parameters that contains the values of the parameters that produce a variation of  $\chi^2_{\nu}$ of 1$\sigma$ from its minimum value.}. We can also calculate the corresponding standard deviation of the mean as an indication of the dispersion of the set of models with $\chi^2_{\nu} \le \chi^2_{\nu, min} + 1.0$\footnote{In doing this, we are assuming that all the SFR(t) that we average in each instance belong to the neighborhood of a single minimum of $\chi^2_{\nu}$}.

\subsection{Tests using input models with known star formation history. Establishing a cut-off value of $\chi^2_{\nu}$}\label{cutoff}

At various stages of this work we have tested the performance of our procedure by trying to retrieve the characteristics of {\it input models} with known SFH. This process has proved to be a very useful tool, for example when delineating the boxes in the CMD, or when defining the intervals of age or the values of the $a_i$ coefficients. Once all this is set, an obvious test is checking how well can we recover a particular SFR(t) with an underlying set of IMF, Z(t) and $\beta(f,q)$, using models computed with different sets of these underlying parameters, and what is the corresponding $\chi^2_{\nu}$ value. We have used two input models with different SFR(t), namely input models A and B (see Figure~\ref{fake_suau}). 

When we try to retrieve the SFR(t) using the same set of underlying parameters, the general trend of the input SFR(t) is successfully recovered, and $\chi^2_{\nu}$ is small ($\chi^2_{\nu} \le $0.7 and 0.5, see Appendix A). When we attempt to recover the SFR(t) of input models A and B by using slightly different combinations of Z(t), IMF and $\beta(f,q)$ with respect to the set used for the input model, we always obtain a ``best fitting'' SFR(t) for each combination of these parameters. But in general, the more different the underlying parameters, the more the output SFR(t) differs from the input SFR(t), and the more the corresponding $\chi^2_{\nu}$ increases. Whether each ``best fitting'' SFR(t) is a good representation of the data is indicated by the value of $\chi^2_{\nu}$ itself. We will see in Section~\ref{sfh} that the $\chi^2_{\nu}$ value associated with each combination of parameters varies substantially, indicating that only some of them are compatible with the data. Here we will discuss the empirical determination of a cutoff value for $\chi^2_{\nu}$.


In calculating $\chi^2_{\nu}$ we implicitly have assumed a Poisson distribution of the number of stars in each box, and applied the error accordingly, as the square root of the number of stars observed in each box. If that were accurate, $\chi^2_{\nu}$  should be close to 1 if the model is a good representation of the data. Nevertheless, the sources of error in the number of stars in each box (incompleteness, S/N, small number statistics, etc) do not necessarily result in Poissonian statistics. For this reason, we used input models A and B to determine the maximum value of $\chi^2_{\nu}$ that still indicates a good representation of the data.  We have tried to recover the two input SFR(t) with a) a synthetic CMD with exactly the same IMF, Z(t) and $\beta(f,q)$ as the input models and b) small differences in these input parameters. These tests are described in Appendix A.   

The results of this experiment indicate the following:

i) values up to $\chi^2_{\nu}\simeq 2.0$ reflect a good agreement between the input and the recovered SFR(t), and indicate also that all the remaining parameters defining the SFH are well represented.

ii) for $\chi^2_{\nu} \gtrsim 3.0$, we obtain results for the SFR(t) that are substantially different from the input SFR(t). The failure in finding the correct SFR(t) is due to the fact that it seems impossible to reproduce the right counts of stars across the CMD if the underlying  Z(t), $\beta(f,q)$, and IMF are not well matched.

iii) differences in the synthetic CMD used to recover the input models as small as $\Delta f$=0.3 in the binary population, $\Delta x$=1.0 in the IMF or small differences in Z(t) produce $\chi^2_{\nu}$ of about 3.0 or more. Therefore, in solutions for the SFH of Leo I with an associated $\chi^2_{\nu}\le 3.0$, such a low value of $\chi^2_{\nu}$ may indicate that the parameters of the model are as close to the actual parameters for Leo I as the differences above, which is remarkable.  

In short we conclude that a limit of $\chi^2_{\nu} \le 2.0$ is a reasonable one for models that are a good representation of the data. We will adopt this limit to determine the best solution for the Leo I SFH.

\section{Results: the Star Formation History of Leo~I} \label{sfh}

Table~\ref{chis} lists the values of $\chi^2_{\nu}$ for the parameter space  discussed in Section~\ref{param}, and assuming a distance modulus $(m-M)_0=22.18$. Table~\ref{tdis} lists the value of $\chi^2_{\nu}$ for the same set of models with $x=-3.2$ in Table~\ref{chis}, but assuming values of the distance modulus in the extremes of the error interval ($\pm 0.1$). All the models in these table have been computed with the mass of the secondary star in the binary systems drawn from the same IMF as the primary star and $0.6\le q \le 1.0$. In an additional set of models covering only a subset of the parameter space represented in Table~\ref{chis}, we have assumed a flat IMF for the secondaries, as suggested by Elson et al. (1998) and Kroupa \& Tout (1992), and different $q$ intervals. Figures~\ref{zeta}, \ref{binaries} and \ref{figimf} show the variation of $\chi^2_{\nu}$ as a function of Z(t), $\beta(f,q)$ and IMF for a set of selected models. The $\chi^2_{\nu}$ listed in Table~\ref{chis}and~\ref{tdis} (and represented in Figures~\ref{zeta} to \ref{figimf}) correspond to the minimum search using the grid method. In an equally systematic search made using a downhill simplex method (Nead \& Mead 1965; Press et al. 1986), we obtained basically the same ordering of the models, form the best to the worst, and similar values of $\chi^2_{\nu}$. The accepted models ($\chi^2_{\nu} \le 2.0$) are highlighted in both Table~\ref{chis} and Figures~\ref{zeta} to~\ref{figimf}. These solutions define a very narrow region of the parameter space characterized by Z=0.0004, a Kroupa et al. IMF or steeper, and a relatively large fraction of binary stars (lower fractions of binary stars are required when we assume a flat IMF for the secondaries). Figures~\ref{perfectox1z0004b3} to~\ref{perfectox1zsigma1b3} show some examples of the best SFR(t) for a number of combinations of the remaining parameters, and the resulting model CMD.    

In the following, we discuss in some detail the conclusions on each of the explored parameters, and the effects on the SFR(t) of varying Z(t), $\beta(f,q)$, and the IMF holding the other parameters fixed. We will also comment on the effects of a shift in the adopted distance modulus.


\subsection{Z(t)} \label{zt}

Metallicity is a parameter that, together with the binary fraction (see below), determines whether we can find models compatible with the data. Figure~\ref{zeta} shows the strong dependence of $\chi^2_{\nu}$ on Z. Very large values of $\chi^2_{\nu}$ are obtained when Z=0.0001 or Z=0.0006. In all cases, the minimum of $\chi^2_{\nu}$, at fixed IMF and $\beta(f,q)$, occurs for Z=0.0004. For this metallicity, the minimum is low enough to provide an acceptable solution ($\chi^2_{\nu} \le 2.0$) in a few instances. In the case of Z increasing with time in the interval Z=0.0001-0.0008, $\chi^2_{\nu}$ is always slightly larger than for Z=0.0004 (except for the IMF with $x=-2.0$, see Table~1) and never compatible with the data using our criteria of $\chi^2_{\nu} \le 2.0$.

Figure~\ref{perfectox1zsigma1b3} shows the synthetic CMD computed with the best solution for the SFR(t) obtained for the IMF with x=--3.2, $f$=0.6 and Z increasing from Z=0.0001 to Z=0.0008. It can be compared with the corresponding best model CMD obtained for the identical IMF and $\beta(f,q)$ but Z=0.0004 in Figure~\ref{perfectox1z0004b3}, and with the observed CMD in Figure~\ref{leoi234}. The RGB of the model with increasing Z(t) is narrower ($\sigma_{V-I, M_I=-2.5} \simeq 0.04$) than the RGB in the observed CMD ($\sigma_{V-I, M_I=-2.5} \simeq 0.07$), whose morphological characteristics are better represented by the RGB obtained for constant Z=0.0004 ($\sigma_{V-I, M_I=-2.5} \simeq 0.05$). Therefore, it turns out that our best model, picked up without using any information on the structure of the RGB, but on the basis of star counts in the CMD alone, is also the one that reproduces best the observed width of the RGB. Note that the change in $\chi^2_{\nu}$ between the models in Figures~\ref{perfectox1z0004b3} and ~\ref{perfectox1zsigma1b3} is small, but interestingly, it is in the same sense as the morphological agreement.

Finally, we tested a set of models with identical parameters as the Z=0.0004, $x=-3.2$ models, but adding a Gaussian dispersion of $\sigma_Z=0.0001$ to the constant Z=0.0004 metallicity. The values of $\chi^2_{\nu}$ indicate worse agreement than in the case of no dispersion in Z. This suggests that not only is Z approximately constant, but also that the dispersion in Z is small.  This prediction of the model needs to be tested by directly measuring the metallicity of a sample of stars. If confirmed, it would point to a very interesting chemical (non) enrichment history for Leo~I, whose detailed knowledge through abundance measurements of individual stars would provide important clues on the unknown processes that regulate the chemical enrichment in dSph galaxies. 

A change in Z(t) produces some changes in the SFR(t) at intermediate and old ages, in the sense that for the lower metallicity (Z=0.0001) there is some star formation at age$\simeq$ 15-12 Gyr and for the larger metallicity (Z=0.0006), there are no stars that old, and the main increase in the SFR(t) takes place around 5 Gyr instead than around 7 Gyr. In short, a lower overall metallicity results in an older stellar population in the mean, while it would be found younger if the metallicity were higher.  This is the usual effect of the much discussed  age-metallicity degeneracy. But it appears that this degeneracy has been broken by our technique, since only Z=0.0004 models provide $\chi^2_{\nu}\le 2.0$, i.e.  are capable of reproducing the star counts across the CMD.


\subsection{$\beta(f,q)$} \label{betafq}

Figure~\ref{binaries} displays the values of $\chi^2_{\nu}$ as a function of the fraction of binaries $f$, for the case of x=--3.2 and Z=0.0004. Panel (a) shows the case of the secondary stars drawn from the same IMF as the primaries, while (b) considers the case of secondaries being drawn from a flat IMF (see Section~\ref{secbin} for a justification of these choices). Different lines show sets of models computed with different $q$ intervals. These results allow us to put some constraints on the characteristics of the binary star population in Leo I.

Since the IMFs we have considered rise steeply towards low-mass stars, when the secondaries are drawn from the same IMF as the primaries, this results in a bias towards binaries with low mass secondaries. A flat IMF for the secondary stars, with some restriction on the mass ratio, tends to produce a larger fraction of similar mass binaries. In the first case, (Figure~\ref{binaries}a), we need a large $f$ to find a model that is a good  representation of the data. If the range of $q$ is limited to $q>0.6$, $f=0.6$ to 0.9 is required. If $q>0.8$, the minimum of $\chi^2_{\nu}$ occurs for $f=0.3$, with $\chi^2_{\nu}=2.4$, which is marginally above our criteria for an accepted model. For $q>0.1$ we don't find any acceptable model. In the case of the secondaries being drawn from a flat IMF, the required fraction of binaries is lower: the minimum of each set of models occurs at $f=0.1$ for $q>0.9$, at $f=0.3$ for $q>0.7$ and $q>0.8$, and at $f=0.6$ for $q>0.6$. However, for $q>0.8$ and $q>0.9$ none of the models is compatible with the data.

From the interdependence between $q$ and $f$ we conclude that, in the context of our models, $f=0.3-0.6$ with $q>0.6$ and approximately flat IMF for the secondaries provides a good fit to the observed CMD of Leo I. Different distribution of $q$ require different $f$ values, but the accepted combinations of $q$ and $f$ produce a similar fraction of near-equal mass binaries than the combination above.  This conclusion is consistent with the results obtained by Elson et al. (1998), Rubenstein \& Baylin (1997) and Kroupa \& Tout (1992) on star clusters.

Changes in $\beta(f,q)$ have very small effects on the SFR(t). With all the remaining parameters fixed, an increase of $f$ tends to slightly decrease the total star formation required to account for the observed CMD, in particular the total star formation at intermediate age. The change of total star formation with $f$ is only $\simeq$ 5\% between $f=0.1$ and $f=0.9$.


\subsection{IMF} \label{imf2} 

Our results are compatible with a Kroupa et al. IMF which, in the mass range more common in our CMDs ($\simeq 1 M_{\odot}$), is very similar to a Salpeter IMF (x=2.2 versus x=2.35), and are slightly better for a steeper IMF ($x=-3.2$). Figure~\ref{figimf} shows the dependence of $\chi^2_{\nu}$ on the slope of the IMF  for the case Z=0.0004. Except for $f=0.1$, $\chi^2_{\nu}$ keeps decreasing from $x=-2.0$ to $x=-3.2$, but this tendency does not continue if we increase the slope further. Unlike Z(t) and $\beta(f,q)$, the dependence of $\chi^2_{\nu}$ on the IMF slope is small and not systematic, and therefore, we have little discrimination on it. The preference for a steeper IMF is consistent with the result of  Holtzmann et al. (1997) for the LMC, where they also find a preference for IMF slopes steeper than the Salpeter value. 

The main effect of changing the IMF is the zero point of the SFR(t), and therefore, the amount of mass in stars and stellar remnants inferred for the galaxy. Most stars in the CMD have masses between 0.5 and 1.5$M_\odot$, and therefore, there is not a strong difference in zero point between the Kroupa et al. (1993) IMF and the IMF with $x=-2.0$, since they have a very similar slope in that mass range. However, their difference with $x=-3.2$ is of about 50\% more mass in the last case. This is because the steeper IMF puts less stars in the CMD per unit of mass of star formation.

\subsection{SFR(t)} \label{sfrt}

The resulting SFR(t) and the corresponding CMD for two accepted models are shown in Figures~\ref{perfectox1z0004b3} and~\ref{bestsfrp2}. Figures~\ref{worstsfr} and \ref{perfectox1zsigma1b3} show two additional models of interest to illustrate the discussion in this section. In particular, Figure~\ref{worstsfr} shows an example of a poor fit of the SFR(t) and the corresponding model CMD. Note that the CMDs shown in these figures are qualitatively very similar to the observed CMD -although some details differ- and, except for Figure~\ref{worstsfr}, almost visually indistinguishable. Although the relative SFR(t) at different $t$ vary slightly from one accepted model to the other, the general trend of the SFR(t) is robust, with its value increasing from 15 to 7 Gyr ago, and most of the star formation taking place between 7 and 1 Gyr ago: for the 5 models with $\chi^2_{\nu}\le 2.0$ (Table~\ref{chis} and Figure~\ref{binaries}), the total fraction of star formation between 7 Gyr ago and 1 Gyr ago ranges from 73\% to 81\%. At 1 Gyr ago, the star formation abruptly drops again to a negligible value, but seems to be active until at least $\simeq 300 $ million years ago. Our results don't unambiguously answer the question of whether Leo I had some star formation starting around 15 Gyr ago, but it seems clear that the amount of this star formation, if it occurred at all, would be small. Stars 15-12 Gyr old would contribute to the bridge of stars observed from the base of the RC to 
the tip of the MS (see Figure~\ref{leoi234}). Stars between 12 and 10 Gyr old are responsible for the little blue extension observed at the base of the RC ($[(V-I)_0, M_I] = [0.65, -0.25]$. These features in the observed CMD seem to indicate indeed some star formation at these old ages.

The resolution in age of our solution has been imposed by the choice of the intervals of age corresponding to the partial models. We obtain an averaged SFR(t) for each of these intervals of time. The error bars in the histograms represent the 1$\sigma$ dispersion around the mean SFR(t) obtained by averaging the models with values of $\chi^2_{\nu} \le \chi^2_{\nu, lim}$ with  $\chi^2_{\nu, lim}= \chi^2_{\nu, min} + 1.0$ (see Section~\ref{minchi}). The relative dispersion of the mean SFR(t) is larger at older ages, indicating that we actually have less information about them in the CMD, both because of the fainter stars involved and the packing of the stars in the CMD. The relatively large error at ages younger than 1 Gyr obtained in some cases may be related to the statistical error associated with the small number of stars at those ages. 

It is also interesting to note that the general trend for the SFR(t) is relatively similar for the models with different sets of IMF, Z(t) and $\beta(f,q)$ that result in large values of $\chi^2_{\nu}$ (see Figure~\ref{worstsfr}). In other words, the result for the SFR(t) is not strongly dependent on the IMF, Z(t) or even $\beta(f,q)$, for reasonable choices of these parameters, but these other parameters are the ones that determine how good the agreement is between the data and the models. 

\subsection{Distance modulus} \label{dist}

Table~\ref{tdis} shows that we don't find any solution for the SFH if we assume a value for the distance modulus at the extremes of its error interval. In addition, the values of $\chi^2_{\nu}$ show relative trends similar to those in Table~\ref{chis}. In most cases, $\chi^2_{\nu}$ increases between 2.5-6.0 units.
Only for Z=0.0001 and $(m-M)_0=22.08$, $\chi^2_{\nu}$ has lower values than for the adopted distance modulus $(m-M)_0=22.18$, but not low enough to produce acceptable models. We only tested the effects of the distance modulus for one of the IMF values ($x=-3.2$). Since the IMF is the parameter that less affects the results, it is reasonable to assume that we would reach similar conclusions for the other IMFs.

This result a) suggests that the adopted value for the distance modulus (at least relative the stellar evolution models used) is correct. This is remarkable, since  the Lee et al. (1993a) TRGB distance is based in the Da Costa \& Armandroff (1990) calibration, who took the slope of the relation between the TRGB brightness and the metallicity from Sweigart \& Gross (1978) models while the zero point was set by empirical $M_{bol}$ values from Frogel et al. 1983, adopting the HB distance scale by Lee et al. 1990  (see also Salaris \& Cassisi 1998); and b) supports further the uniqueness of our solution (Section~\ref{unicity}), since it implies that no combination of the model parameters is compensating uncertainties in the distance.    

\subsection{Uniqueness of the solution} \label{unicity}

Using the criterion that $\chi^2_{\nu} \le 2.0$ results in three acceptable models amongst those presented in Table~1, plus some additional models compatible with the data in the additional tests with different binary population characteristics. No solution has been found for values of the distance modulus at the extremes of its error interval (Table~2). Obviously, there would be many other solutions if we divided the space of parameters more finely. Strictly speaking, then, our solution is not unique. Note, however, that these solutions define a narrow range of acceptable Z(t), $\beta(f,q)$ and IMF, i.e. define a minimum of $\chi^2_{\nu}$ in a well defined position of the parameter space: 

$\bullet$ Z(t)=0.0004 $\pm$ 0.0001, 

$\bullet$ Kroupa et al. (1993) IMF or steeper up to $x=-3.2$, $0.5\le m < 30$,  

$\bullet$ $\beta(f,q)$ with $f=0.3-0.6$ and $q>0.6$ and approximately flat IMF, or particular combinations of these parameters that would produce a similar fraction of almost equal mass ($q>0.7$) binaries.

$\bullet$  $(m-M)_0=22.18$ with $\Delta (m-M) < 0.1$.


One may question whether there are sets of correlated parameters that act in such a way that suitable changes in all of them would compensate, producing various sets of acceptable solutions. Except for the case of the pair ($f$, $q$) discussed above, from Table~1 and 2 it appears that this is not the case. In particular, the fact that we only find solutions for a particular value of the distance modulus strongly supports an uniqueness of the solution. The same conclusion is suggested by the fact that, in the tests with input models A and B (Appendix A), relatively small changes in the assumed Z(t), $\beta(f,q)$ and IMF between the input models and the synthetic CMD, led to unacceptable solutions for the SFR(t).

\section{Summary and final remarks}

We have presented a detailed study of the SFH of the Local Group dSph galaxy Leo I. We have adapted the method presented in Aparicio et al. (1997b) to the case of data reaching as deep as the oldest MS turnoffs. We have explored a wide range of values for the metallicity Z(t), the binary fraction $\beta(f,q)$ and the IMF, and for each combination of these parameters, $\simeq 6 \times 10^7$ possible SFR(t) have been tested. The results of this analysis allow us to confidently constrain this space of parameters defining the SFH. Our solution for the SFH is bounded within some range of possible input parameters, since it defines a minimum of $\chi^2_{\nu}$ in a well defined position of the parameter space. 

Our results for the SFR(t) are robust in the sense that the main characteristics of the SFR(t) are unchanged for different combinations of the remaining parameters. However, only a narrow range of assumptions for Z(t), IMF and $\beta(f,q)$ result in a good agreement between the data and the models, namely: Z=0.0004, a Kroupa et al. IMF or slightly steeper, and a relatively large fraction of binary stars. Most star formation activity (73\% to 81\%) occurred between 7 and 1 Gyr ago. At 1 Gyr ago, it abruptly dropped to a negligible value, but seems to have been active until at least $\simeq 300 $ million years ago. Our results don't unambiguously answer the question of whether Leo I began forming stars around 15 Gyr ago, but it appears that the amount of this star formation, if existing at all, would be small. Stars 15-12 Gyr old would contribute to the bridge of stars observed from the base of the RC to the tip of the MS, and stars between 12 and 10 Gyr old would be responsible for the little extension to the blue observed at the base of the RC. The presence of these features in the observed CMD seem to indeed indicate some star formation at old ages. In Figure~\ref{fincolores} we represent one of the accepted model CMD using different colors for each partial model. This provides explicit information of the age interval that populates each region of the CMD. 


Metallicity and binary stars are the two parameters, together with the SFR(t), that determine whether a model that is a good representation of the data can be found. Our conclusions on both of them are supported by independent evidence:    

i) the best Z(t), picked up on the basis of star counts in the CMD alone, is also the one that produces the best morphological agreement with characteristics of the data like the width of the RGB (Figures~\ref{perfectox1z0004b3} and \ref{perfectox1zsigma1b3}). The conclusion, however, is intriguing: it implies that little or no metallicity evolution has occurred in Leo I after an early epoch, and therefore, it provides the testable prediction that a small metallicity dispersion should be found among Leo I stars. With this aim, we plan to obtain Ca II triplet spectroscopy of stars in Leo I using the VLT. If this prediction is confirmed, it would point to a very interesting chemical (non) enrichment history for Leo~I, whose detailed knowledge through abundance measurements of individual stars, would provide important clues on the unknown processes that regulate the chemistry of dSph galaxies.  

ii) our conclusion about the characteristics of the binary star population, parameterized by $\beta(f,q)$, is consistent with the results obtained by Elson et al. (1998), Rubenstein \& Bailyn (1997) and Kroupa \& Tout (1992) on star clusters, and show the relevance of this parameter, which has been usually neglected in SFH studies.

Finally, the fact that we don't find any solution for values of $(m-M)_0$ at the extremes of its error interval supports the adopted distance modulus and the uniqueness of our solution, since it implies that no combination of the model parameters is compensating distance errors.
 
 We provide a quantitative measure of the ``goodness of fit'' through the value of the chi-squared statistic. The experiments are ``calibrated'' by using input models with known SFHs. The fact that the best $\chi^2_{\nu}$ obtained when comparing models and data is similar to the one resulting when retrieving the SFH of input models using models with similar Z(t), IMF and $\beta(f,q)$ suggest that we are indeed constraining these parameters for Leo I. Finally, the fact that it is possible to find solutions (i.e. low enough values of $\chi^2_{\nu}$) is a success for stellar evolution models. 
\acknowledgments

We want to thank R. Carlberg, M. Easterbrook, J. Harris, R. Marzke, S. Mauch and F. Pont for very useful discussions about the $\chi^2_\nu$ minimization. Thanks are also due to almost everyone at Carnegie Observatories for allowing us to intensively use the CPUs of their workstations during the never-ending artificial star tests or the searches for the minimum in the $\chi^2_\nu$ space. Support for this work was provided by NASA grant GO-5350-03-93A from the Space Telescope Science Institute, which is operated by the Association of Universities for Research in Astronomy Inc. under NASA contract NASA5--26555.  C.G. also acknowledges financial support from a Small Research Grant from NASA administered by the AAS and a Theodore Dunham Jr. Grant for Research in Astronomy. A.A. is supported by the Ministry of Education and Culture of the Kingdom of Spain and by the IAC (grant PB3/94). 

\appendix
    
\section{Testing the method by recovering input models with known star formation histories}

In several stages of this work we have used ``input models'' with known SFHs to test our level of success in retrieving the SFH. This process has proved to be a very useful tool, for example when delineating the boxes in the CMD, or when defining the intervals in age or the values of the $a_i$ coefficients. In this Appendix we will concentrate on the use of these tests to establish the range of values of $\chi^2_{\nu}$ that indicate agreement between the model and the data.  

$\chi^2_{\nu}$ is a statistic that characterizes the dispersion of the observed frequencies from the expected frequencies. The numerator in equation (2) in Section~\ref{method} is a measure of the spread of the observations with respect to the model. The denominator is the expected spread assuming Poisson statistics. For a model representing the data, the average spread of the data should correspond to the expected spread, and thus we would get a contribution of about 1 for each degree of freedom. Since we have imposed the constraint that the total number of observed stars in the boxes must be equal to the number of model stars in them, the number of degrees of freedom in our experiment is $\nu= r - 1$, where $r$ is the number of boxes. In theory, the expected value of $\chi^2_{\nu}$ for a good model should be $\chi^2_{\nu} \simeq 1$, for which the probability associated will be approximately 0.5 (Bevington \& Robinson 1992). Nevertheless, to apply the test in this way, it is necessary to have a good understanding of the errors. We have assumed a Poisson distribution of the number of stars in each box, and estimated the variance accordingly. However, the sources of error in the number of stars in each box (crowding, S/N, small number statistics) do not necessarily result in Poissonian statistics.  

Here, we have used input models with known SFH to determine the value of $\chi^2_{\nu}$ that should be expected in the case of a good model, and to decide what cutoff value of $\chi^2_{\nu}$ indicates that the model is no longer a good representation of the data. We have done this by trying to recover the (known) SFR(t) of the input models with a set of partial models computed with the same IMF, Z(t) and $\beta(f,q)$. In this way, the difference between the original and the recovered SFR(t) is basically due to the statistical uncertainties and to the intrinsic limitations of the method (like, for example, the fact that the $a_i$ have fixed values in finite steps of 0.2). In successive tests, we attempted to recover the input SFR(t) by using a set of partial models that differed from the input model in either Z(t), IMF, $\beta(f,q)$ or in a small difference (within the error interval) in distance modulus (see Table~\ref{tests}). 

For all these tests we have used two input models, that we will call input model A  and input model B, with the same Z(t), IMF, and $\beta(f,q)$  (Z(t) increasing with time from Z=0.0001 to Z=0.0004,  Kroupa et al. IMF and no binary stars), but different SFR(t), as shown in Figure~\ref{fake_suau}. 


Figure~\ref{outfake} show the comparison between the input SFR(t) and the output SFR(t) for the corresponding cases in Table~\ref{tests}. The values of $\chi^2_{\nu}$ corresponding to the solution for the identical set of parameters are 0.7 and 0.5 for input models A and B respectively. The fact that they are smaller than 1 may indicate that the assumption of Poisson statistics in the number of stars in the CMD boxes actually results in an overestimate of the errors. When the difference in the input parameters is $\Delta f=0.1$ or a small difference in the IMF, $\chi^2_{\nu}$ increases to about $\simeq 1.0$, and the shape of the SFR(t) is still quite well recovered. For larger differences in the IMF or $\beta(f,q)$,  for differences in Z(t) or for a difference in the distance within the error interval ($\Delta (m-M)_0=\pm0.1$), $\chi^2_{\nu} \ge 3.0$ and the recovered SFR(t) is substantially different from the input value. 


This simple experiment shows that:

i)  values up to  $\chi^2_{\nu} \simeq 2.0$ reflect a good agreement between the input and the recovered SFR(t), and indicate also that all the remaining parameters defining the SFH are well represented.

ii) for  $\chi^2_{\nu} \ge 3.0$ we obtain results on the SFR(t) that are substantially different from the input SFR(t). The origin of the failure to find the `correct' SFR(t) seems to be that it is impossible to produce the right counts of stars in the CMD if {\it underlying} parameters like the IMF, $\beta(f,q)$ or Z(t) are not well matched, or if other parameters like the distance are incorrect.

iii) differences in the synthetic CMD used to recover the input models as small as $\Delta f$=0.3 in the binary population, $\Delta x$=1.0 in the IMF, $\Delta (m-M)_0 =\pm 1$, or small differences in Z(t) produce $\chi^2_{\nu} \ge 3.0$. Therefore, since we find solutions for the SFH of Leo I with an associated $\chi^2_{\nu}\le 3.0$, it indicates that the parameters of the model are as close to the actual parameters for Leo I as these differences, which is remarkable.  We have not tested, for example, how small variations in the physics of the stellar evolutionary models would affect the value of $\chi^2_{\nu}$. However, the good agreement between the observed and model CMDs based on the Padova stellar evolution models is encouraging.

\newpage

\figcaption[Gallart.fig1.eps]{Observed CMD of Leo~I from the observations obtained with the three WF chips. A distance modulus $(m-M)_0=22.18$ (Lee et al. 1993a) and reddening E(B-V)=0.02 (Burstein \& Heiles 1984) have been used to transform to absolute magnitudes and unreddened colors.\label{leoi234}}
 

\figcaption[Gallart.fig2col.eps]{Set of partial models for a synthetic CMD computed with constant SFR(t) from 15 Gyr ago to the present, Z=0.0004, Kroupa et al. (1993) IMF, and no binary stars. The left panel shows the theoretical synthetic CMD while in the CMD of the right panel, observational errors have been simulated (see Appendix~A). Note the sequence of ages in both the MS and the subgiant-branch, and although less definite, also in the RC and HB.\label{colores}}

\figcaption[Gallart.fig3.eps]{Location of the 18 regions defined to compare the star counts in the observed and model CMDs.\label{cajas6}}

\figcaption[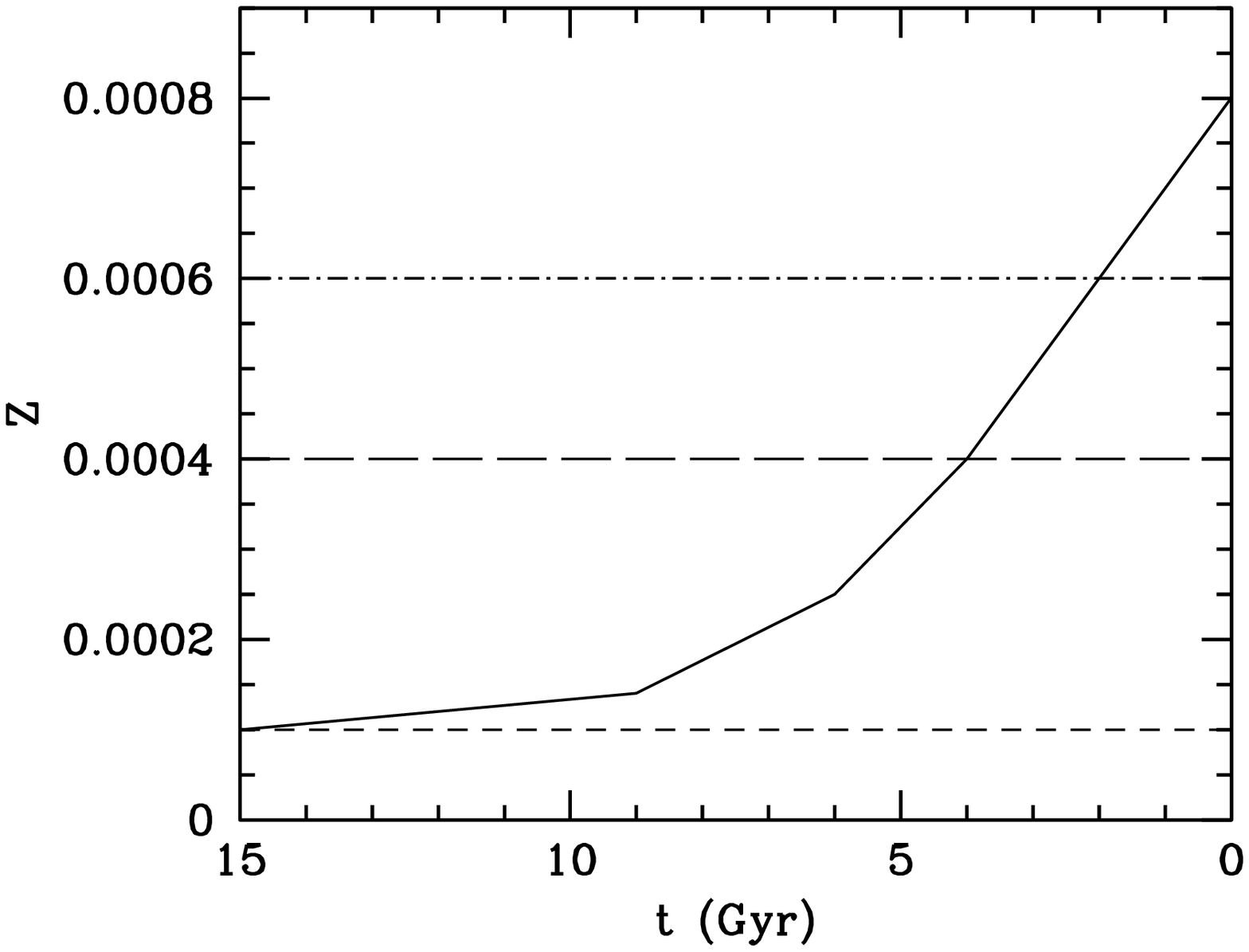]{The four chemical enrichment laws considered in the analysis of the SFH of Leo I. In the case of Z(t) increasing with time, a Gaussian dispersion of Z with $\sigma_Z=0.0001$ has been added at each age.\label{zlaws}}

\figcaption[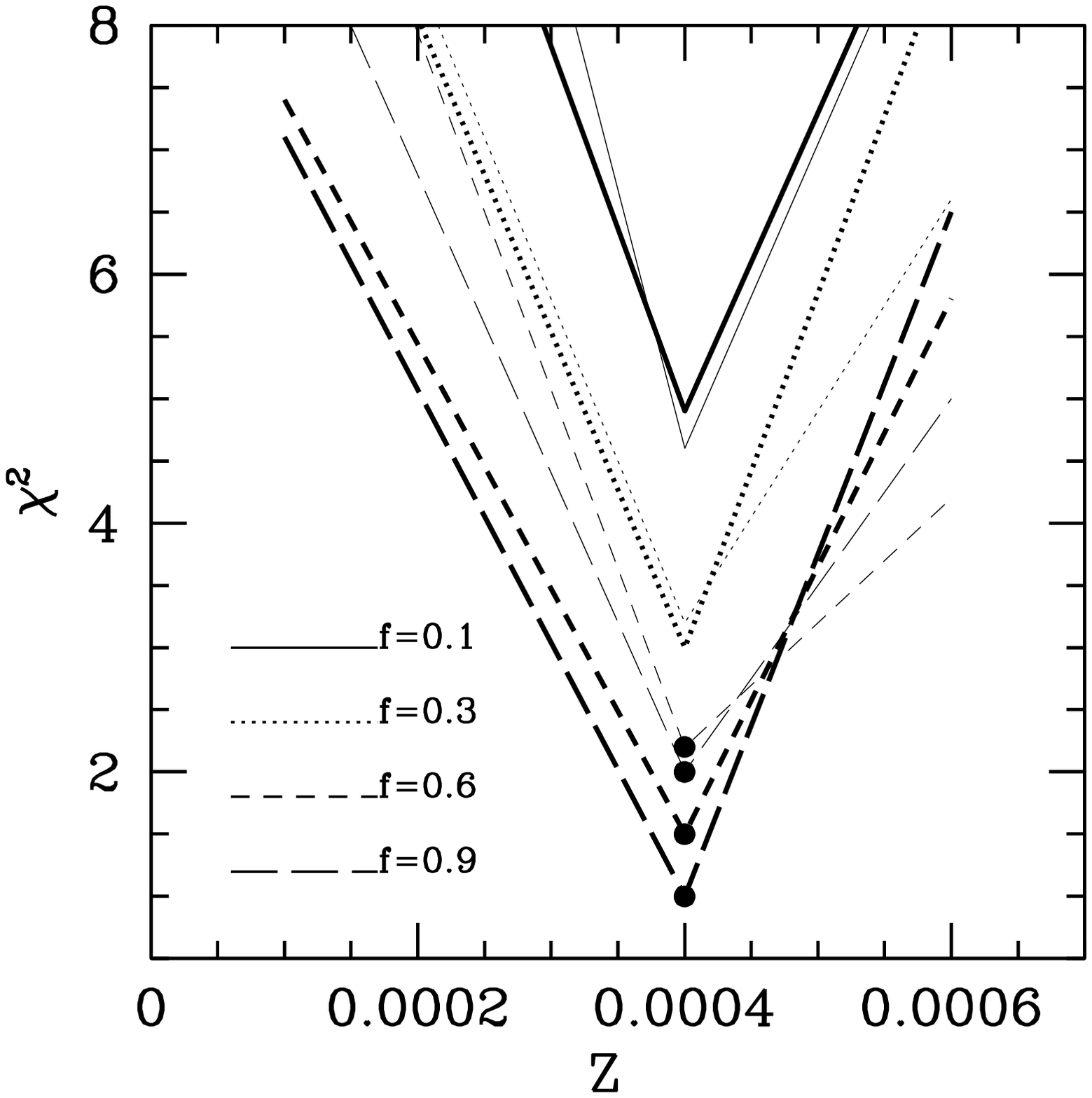]{Dependence of $\chi^2_{\nu}$ on metallicity. Thick lines refer to the models with $x=-3.2$ and thin lines to those with the Kroupa et al. IMF. Accepted models are indicated with solid circles.\label{zeta}}

\figcaption[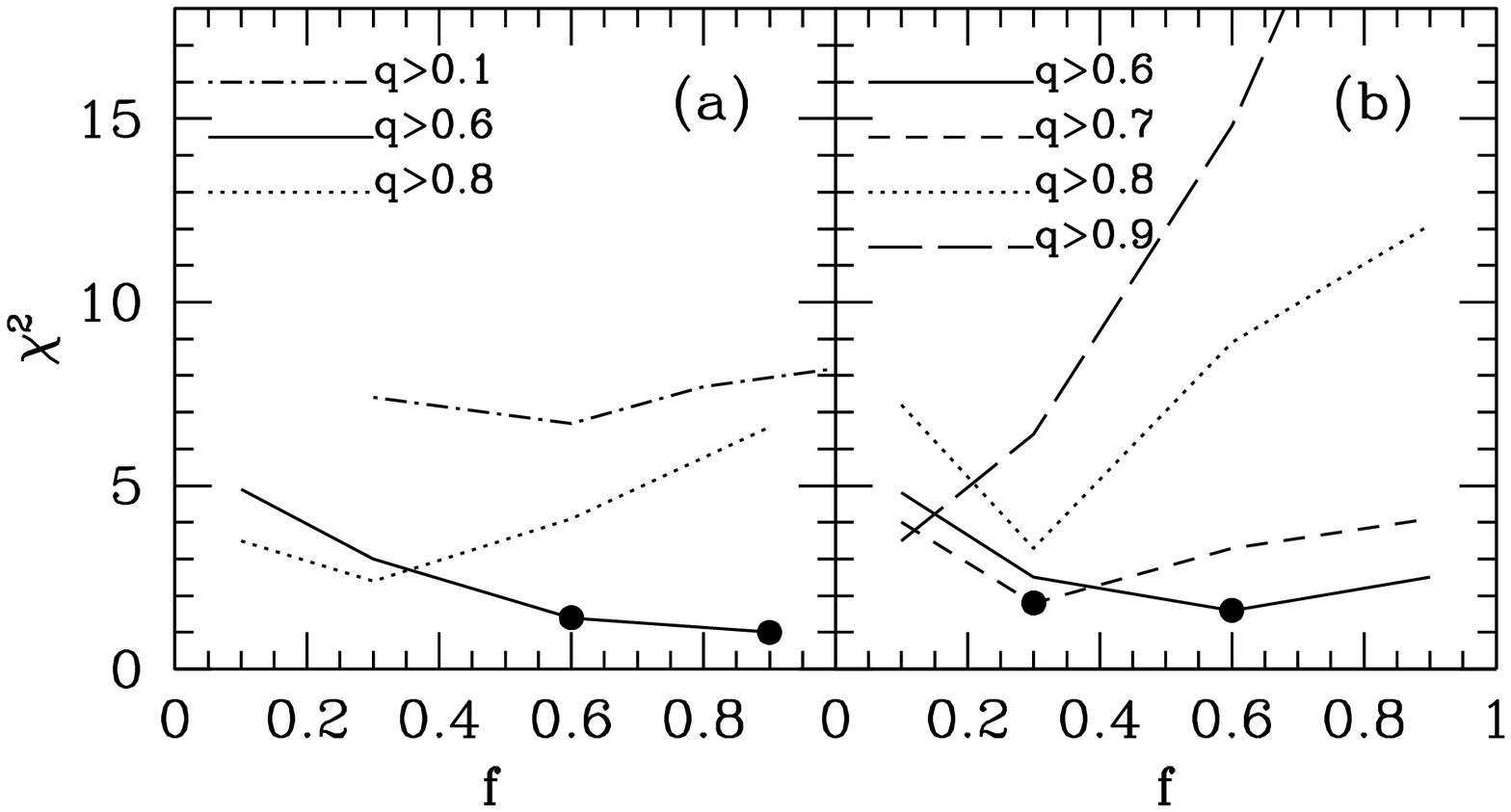]{Dependence of $\chi^2_{\nu}$ on $f$ and $q$. Models in panel (a) have been computed with a flat IMF for the secondaries in binary systems (for $q \ge 0.1$ different $f$ values have been tested, from $f=0.3$ to $f=1.0$). Those in panel (b) have been computed with identical IMF for both stars of the binary systems. Accepted models are indicated with solid circles.\label{binaries}}

\figcaption[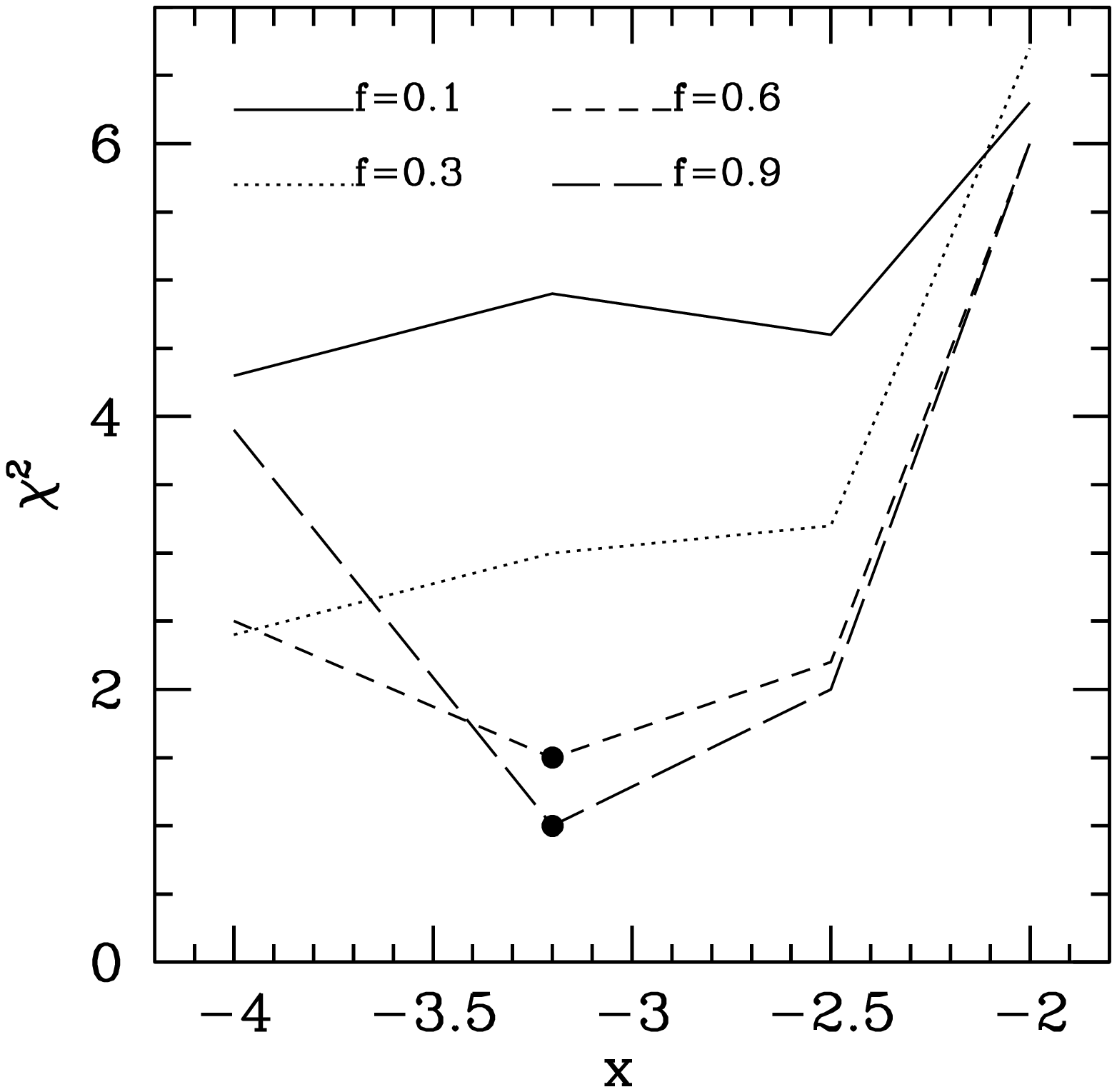]{Dependence of $\chi^2_{\nu}$ on IMF. All models have Z=0.0004. The different line types indicate different $f$, as indicated in the labels. Accepted models are indicated with solid circles.\label{figimf}}

\figcaption[Gallart.fig8.eps]{Upper panel: best SFR(t) obtained for a model with Z=0.0004, IMF with $x=-3.2$ (see Section~\ref{imf1}), and $f=0.6$ with $q \ge 0.6$. Lower panel: corresponding model CMD.\label{perfectox1z0004b3}}

\figcaption[Gallart.fig9.eps]{Upper panel: best SFR(t) obtained for a model with Z=0.0004, $x=-3.2$ IMF (see Section~\ref{imf1}), and $f=0.3$ with $q \ge 0.7$ and flat IMF for the secondaries. Lower panel: corresponding model CMD.\label{bestsfrp2}}

\figcaption[Gallart.fig10.eps]{Upper panel: example of a bad fit SFR(t), for a model with Z=0.0001, IMF with $x=-3.2$ (see Section~\ref{imf1}), and $f=0.1$ with $q \ge 0.6$. Lower panel: corresponding model CMD.\label{worstsfr}}

\figcaption[Gallart.fig11.eps]{Upper panel: best SFR(t) obtained for a model with Z=0.0001-0.0008, IMF with $x=-3.2$ (see Section~\ref{imf1}), and $f=0.6$ with $ q \ge 0.6$. Lower panel: corresponding model CMD. Note the difference in width near the tip of the RGB with respect to the CMD in Figure~\ref{perfectox1z0004b3} (see text for details).\label{perfectox1zsigma1b3}}

\figcaption[Gallart.fig12col.eps]{One of the accepted CMDs. Different colors represent stars in different age intervals.\label{fincolores}}

\figcaption[Gallart.fig13.eps]{SFR(t) and CMD of input models A and B, used for the tests presented in Appendix A.\label{fake_suau}}

\figcaption[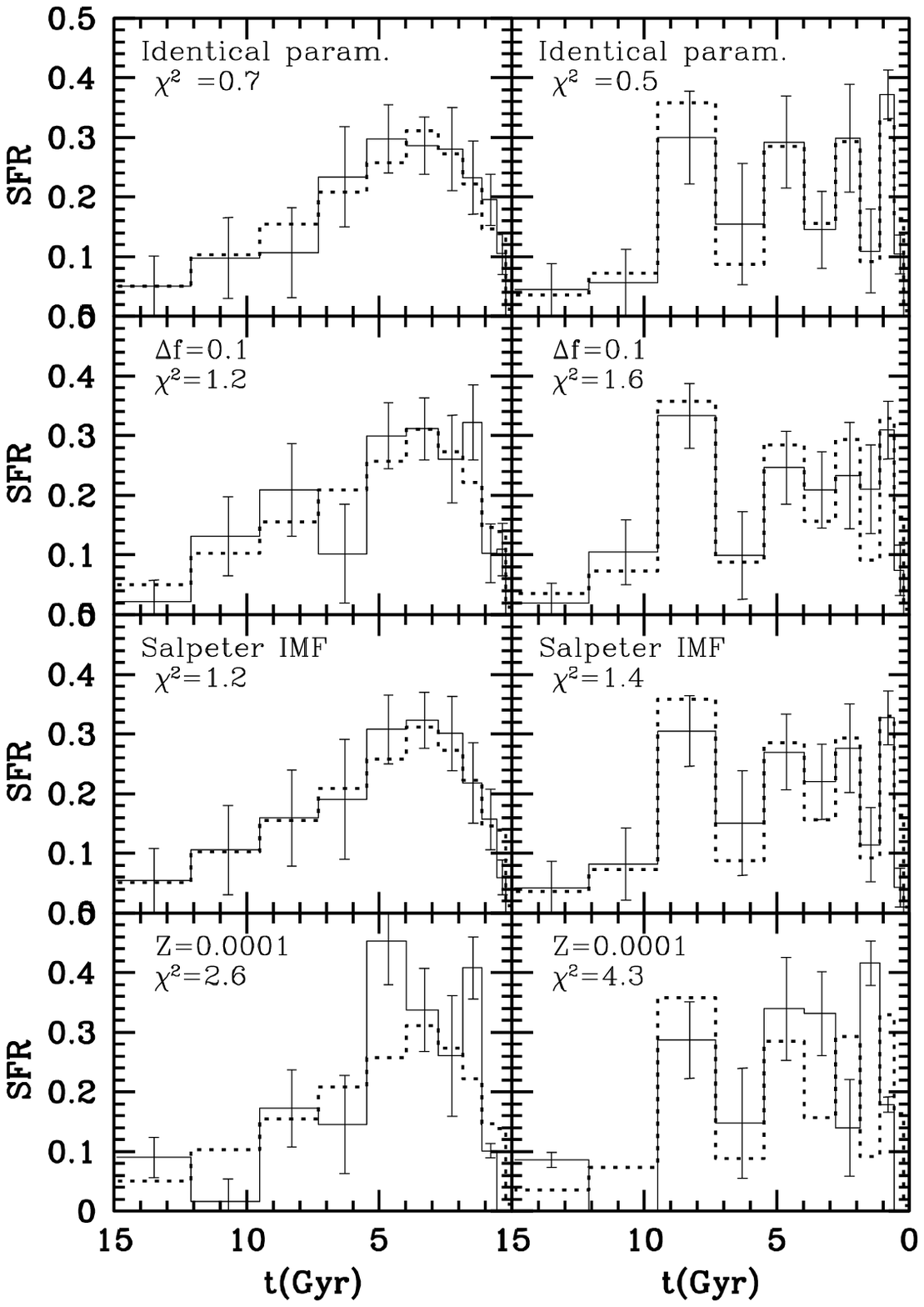]{Comparison between the input SFR(t) and the output SFR(t) in the following situations, from top to bottom: a)the model used has identical parameters as the input model; b) as a), except for the fraction of binaries which is 10\% in the output, versus 0\% in the input; c) as a) except for the IMF which is Salpeter in the input versus Kroupa in the output ($x=-2.35$ and $x=-2.7$ respectively); and d) as a), except for Z(t), which is Z=0.0001 in the output versus Z=0.0001-0.0008 in the input. Note how the input and output SFR become more different as $\chi^2_{\nu}$ increases.\label{outfake}}

\newpage

\begin{table}
\caption {$\chi^2_{\nu}$ values for the best model SFR  for each combination of IMF, Z(t) \& $\beta(f,q)$. A distance modulus $(m-M)_0=22.18$ has been adopted.} 
\begin{center}
\begin{tabular}{r|rrrr}
\hline
\hline
\noalign{\vspace{0.1 truecm}}
Z(t) $\rightarrow$ & Z=.0001 & Z=.0004 & Z=.0006 & Z=.0001-8 \\
\noalign{\vspace{0.1 truecm}}
\hline
\hline
\noalign{\vspace{0.1 truecm}}
$f$ $\downarrow$ & \multicolumn{4}{c}{IMF $x=-3.2$}\\
\noalign{\vspace{0.1 truecm}}
\hline
\noalign{\vspace{0.1 truecm}}
0.1 & 13.7  & 4.9 & 9.7 & 5.2\\
0.3 & 10.6  & 3.0 & 7.7 & 3.6\\
0.6 &  7.4  & \fbox{1.5} & 5.8 & 2.7\\
0.9 &  7.1  & \fbox{1.0} & 6.5 & 2.8\\
\noalign{\vspace{0.1 truecm}}
\hline
\hline
\noalign{\vspace{0.1 truecm}}
& \multicolumn{4}{c}{IMF Kroupa et al.}\\
\noalign{\vspace{0.1 truecm}}
\hline
\hline 
\noalign{\vspace{0.1 truecm}}
0.0 & 19.6 & 7.8 & 12.9 & 7.7\\
0.1 & 17.1 & 4.6 &  9.5 & 5.6\\
0.3 & 11.0 & 3.2 &  6.6 & 3.3\\
0.6 & 10.8 & 2.2 &  4.2 & 3.3\\
0.9 &  9.2 &  \frame{\framebox{2.0}} &  5.0 & 4.6\\
\noalign{\vspace{0.1 truecm}}
\hline
\hline
\noalign{\vspace{0.1 truecm}}
 & \multicolumn{4}{c}{IMF $x=-2.0$}\\
\noalign{\vspace{0.1 truecm}}
\hline
\hline 
\noalign{\vspace{0.1 truecm}}
0.1 & 19.7 & 6.3 & 10.9 & 5.9\\
0.3 & 16.8 & 6.7 & 7.7 & 4.6\\
0.6 & 10.9 & 6.0 & 6.2 & 4.2\\
0.9 & 12.4 & 6.0 & 8.0 & 5.3\\
\noalign{\vspace{0.1 truecm}}
\hline
\hline 
\end{tabular}
\label{chis}
\end{center}
\end{table}

\newpage

\begin{table}
\caption {$\chi^2_{\nu}$ values for the best model SFR with IMF with $x=-3.2$ and combinations of Z(t) \& $\beta(f,q)$ in the case of adopting values for the distance modulus at the extreme of its error interval.} 
\begin{center}
\begin{tabular}{r|rrrr}
\hline
\hline
\noalign{\vspace{0.1 truecm}}
Z(t) $\rightarrow$ & Z=.0001 & Z=.0004 & Z=.0006 & Z=.0001-8 \\
\noalign{\vspace{0.1 truecm}}
\hline
\hline
\noalign{\vspace{0.1 truecm}}
$f$ $\downarrow$ & \multicolumn{4}{c}{$(m-M)_0=22.08$}\\
\noalign{\vspace{0.1 truecm}}
\hline
\noalign{\vspace{0.1 truecm}}
0.1 & 8.3  & 6.9 & 12.1 & 5.3\\
0.3 & 6.5  & 4.7 & 9.8 & 3.8\\
0.6 & 4.2  & 3.7 & 8.7 & 2.7\\
0.9 & 3.9  & 3.8 & 9.6 & 2.5\\
\noalign{\vspace{0.1 truecm}}
\hline
\hline
\noalign{\vspace{0.1 truecm}}
& \multicolumn{4}{c}{$(m-M)_0=22.28$}\\
\noalign{\vspace{0.1 truecm}}
\hline
\hline 
\noalign{\vspace{0.1 truecm}}
0.1 & 19.7 & 9.3 &  12.5 & 10.0\\
0.3 & 16.6 & 6.9 &  11.2 & 8.5\\
0.6 & 13.1 & 5.6 &  8.2 & 7.2\\
0.9 & 12.7 & 4.3 &  8.9 & 7.7\\
\noalign{\vspace{0.1 truecm}}
\hline
\hline
\end{tabular}
\label{tdis}
\end{center}
\end{table}

\newpage

\begin{table}
\caption {Values of $\chi^2_{\nu}$ obtained in the analysis of the input models A and B, for different sets of Z(t), IMF and $\beta(f,q)$. As a reference, both input models are computed with Z(t) increasing from Z=0.0001 to Z=0.0004, Kroupa et al. IMF and no binary stars).}
\label{tests} 
\begin{center}
\begin{tabular}{lccc}
\hline
\hline
\noalign{\vspace{0.1 truecm}}
Parameter changed & Input model A & Input model B \\
\noalign{\vspace{0.1 truecm}}
\hline
\noalign{\vspace{0.1 truecm}}
Identical parameters & 0.5   & 0.3  \\ 
Different age intervals & 0.4 & 0.7  \\
$f=0.1$ &  0.9  & 1.2  \\
$f=0.3$ &  3.3  & 3.8  \\
Salpeter IMF &  1.1 & 1.0\\ 
IMF $x=-3.2$ & 1.2 & 1.2 \\
Z=0.0001 & 2.1& 3.7\\
Z=0.0006 & 17.2 & 15.2\\
$\Delta$ (m-M)=0.1 & 3.5 & 2.8\\
\noalign{\vspace{0.1 truecm}}
\hline
\hline 
\end{tabular}
\end{center}
\end{table}

\end{document}